\documentclass[
 reprint,
 amsmath,amssymb,
 aps,
 prapplied,
 footinbib,
 floatfix
]{revtex4-2}
\usepackage{xcolor}
\usepackage{graphicx}
\usepackage{dcolumn}
\usepackage{bm}
\usepackage{appendix}
\usepackage{chngcntr}
\usepackage{tikz}
\usetikzlibrary{calc} 
\usetikzlibrary{matrix}
\usepackage{float}

\begin{document}

\title{Quantum-Device Simulation of Optical Decoherence of Hole-Spin Qubits in Self-Assembled Quantum Dots}



\author{Jyun-Jie Jiang\textsuperscript{1}, Pericles Philippopoulos\textsuperscript{2}, Félix Beaudoin\textsuperscript{2}, Hong Guo\textsuperscript{1}}

\affiliation{\textsuperscript{1}Centre for the Physics of Materials and Department of Physics, McGill University, Montréal, Quebec H3A 2T8, Canada}
\affiliation{\textsuperscript{2}Nanoacademic Technologies Inc., Suite 1603, 666  Sherbrooke Street West, Montreal, Quebec H3A 1E7, Canada}

\begin{abstract}
Spin--photon interfaces are essential for communications between distant spin qubits in quantum technologies, but the interband optical excitation can also damp electrically driven hole-spin Rabi oscillations in semiconductor self-assembled quantum dots (SAQDs). We report a device-level modeling workflow that integrates realistic SAQD geometry and multiband electronic-structure analysis with models of electrically driven spin control, interband optical transitions, and open-system dynamics. This workflow enables device-level estimation of Rabi-oscillation damping arising from repeated interband absorption--emission cycles. As an example, for a gated GaAs SAQD subjected to a uniform magnetic field $B_0$ along the growth direction of the SAQD, we predict the Rabi frequency of the hole spin qubit and its damping under external illumination. At $B_0=2~\mathrm{T}$, the calculations yield a hole-spin Rabi frequency of 37.3$~\mathrm{MHz}$. When the electrically driven SAQD is illuminated by a broadband LED centered at a wavelength of 790$~\mathrm{nm}$, increasing the optical power from 0.3 to 1.5$~\mathrm{mW}$ shortens the Rabi-oscillation decay time from 90.3 to 17.5$~\mathrm{ns}$. Increasing the SAQD height reduces the electron--hole overlap and thus the emission rate, but the resulting redshift moves the interband transitions into stronger spectral overlap with the LED spectrum, thereby increasing the rate of repeated absorption--emission cycles and enhancing photon-induced Rabi-oscillation damping. The results show that geometry, spin-control conditions, and illumination spectrum should be co-optimized in semiconductor spin--photon devices.
\end{abstract}
\maketitle

\section{Introduction}

Semiconductor spin qubits are promising platforms for quantum information processing because they combine long coherence times, compatibility with established semiconductor manufacturing technologies, and the potential for large-scale integration~\cite{A_crossbar,Qubits_made, An_electrically_driven, Coherent_Single_Electron}. Significant progress has been achieved in spin-based quantum computing using gate-defined quantum dots (QDs)~\cite{Single-shot_read-out,Electrically_driven, Analysis_and3D_TCAD, A_CMOS_silicon_spin}, donor impurities~\cite{ A_silicon-based_nuclear_spin}, and self-assembled semiconductor nanostructures. Among these architectures, self-assembled quantum dots (SAQDs)~\cite{Excitonic_lifetimes,Highly_uniform} offer unique opportunities for realizing hybrid spin--photon quantum systems due to their strong interband optical transitions, nanoscale confinement, and compatibility with optical initialization, manipulation, and readout schemes~\cite{Optically_programmable}. These properties make SAQDs attractive candidates for quantum repeaters~\cite{Optoelectronic}, distributed quantum computing~\cite{Quantum_computation, Universal_quantum,Operation_of}, quantum networking~\cite{Conditional_teleportation}, and spin--photon transduction applications~\cite{Entanglement_distribution}.

The development of scalable quantum technologies increasingly requires reliable interfaces between stationary qubits and flying photonic qubits~\cite{Observation_of_entanglement, Coherent_Transfer_of_Light}. Optical control enables fast qubit initialization, spin manipulation, and quantum-state transfer, while photonic channels provide a natural mechanism for long-distance quantum communication~\cite{Quantum_dot_spin_photon}. However, coupling spin qubits to optical fields introduces additional decoherence pathways~\cite{Decoherence}. Optical excitation can modify spin dynamics through radiative transitions, nonresonant excitation processes, and photon-assisted relaxation channels, which are detrimental to qubit performance~\cite{QD_Optical_Spin_Locking, Direct, Coherent_Optical_Control_of_a, Fundamental_limits}. Understanding and quantitatively predicting this effect is essential for the design and optimization of spin--photon quantum hardware. This work quantitatively examines how repeated interband absorption--emission cycles damp electrically driven hole-spin Rabi oscillations.
\begin{figure*}
\includegraphics[width=1\textwidth]{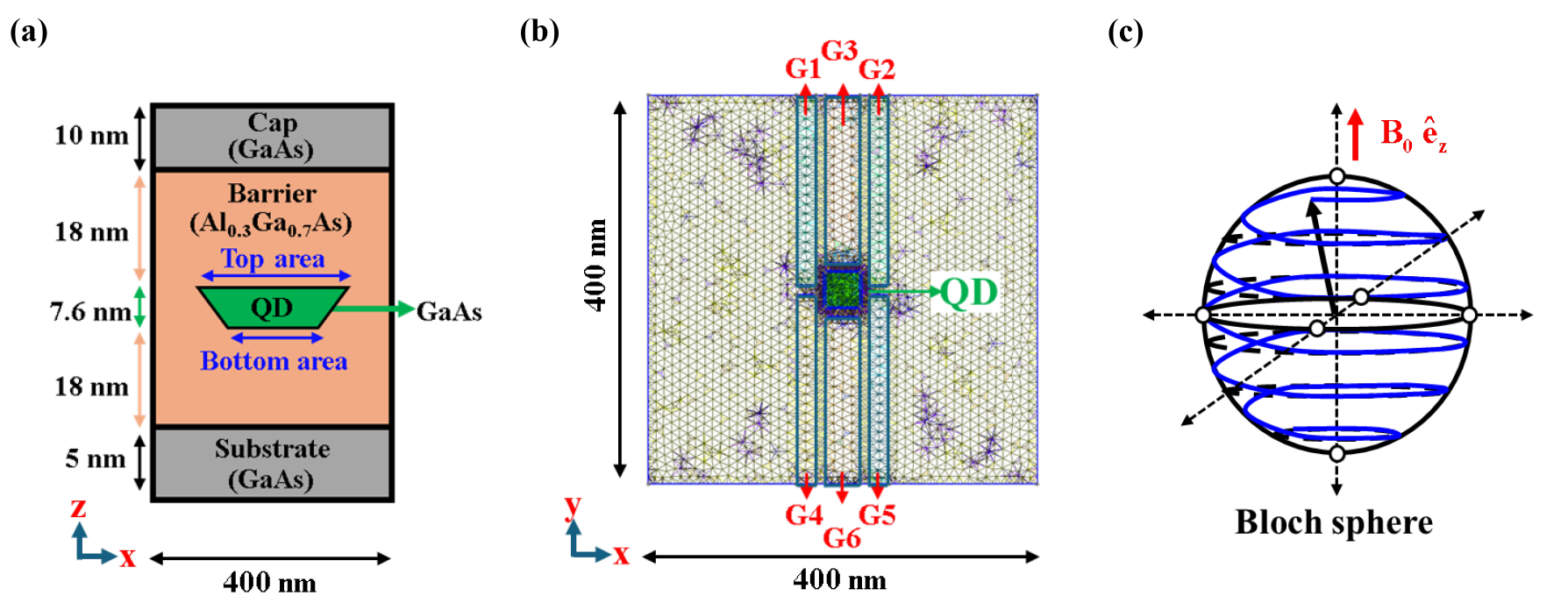}
\caption{\label{fig:Figure_1} (a) Side view of the truncated-pyramidal GaAs SAQD with top and bottom areas of 1253$~\mathrm{nm}^2$ and 400$~\mathrm{nm}^2$, respectively, following the experimental device of Ref.~\cite{Excitonic_lifetimes}. (b) Top view of the GaAs SAQD with six metal gates labeled G1--G6, which are used to tune the energy levels of the SAQD and drive EDSR when an oscillating voltage is applied to gate G6. (c) Rabi oscillations of the spin qubit on the Bloch sphere, where $\mathbf{B}=(0,0,B_0)$ is a uniform and static external magnetic field.}
\end{figure*}

To quantitatively predict this behavior, a unified device-level simulation framework is required. Experimental investigations have demonstrated optical control and spin--photon coupling in a variety of semiconductor QD systems~\cite{Coherent_Optical,Quantum_dot_spin_photon}, including InGaAs and GaAs SAQDs. At the same time, significant theoretical efforts have focused on describing the electronic structure, spin manipulation, and optical transitions in confined semiconductor nanostructures~\cite{Spin_state_tomography, Coherent_Transfer_of_Light, Optical_Orientation}. Electronic states are commonly described using multiband $k \cdot p$ approaches~\cite{Motion_of_Electrons,Quantum_Theory_of_Cyclotron,For93}, atomistic tight-binding methods~\cite{Atomistic}, or effective-mass approximations~\cite{Self-consistent, Multiband_theory}. Spin manipulation is commonly investigated through electric-dipole spin resonance (EDSR)~\cite{Electrically_driven,Electric-dipole-induced} and electron spin resonance (ESR)~\cite{Driven_coherent}, while dissipative spin dynamics are described using open-quantum-system approaches ~\cite{Open_quantum_systems}. In theoretical studies, electronic structure, spin control, optical excitation, and open-system dynamics are often treated separately. This separation limits device-level predictions of how device geometry and operating conditions jointly determine photon-induced damping of electrically driven hole-spin Rabi oscillations.

To address this need, we develop a predictive quantum-device simulation workflow for gated SAQD spin qubits that connects realistic device geometry and multiband electronic structure to electrically driven hole-spin dynamics and photon-induced Rabi-oscillation damping. The electronic structure is calculated using a finite-element implementation~\cite{Robust} of the multiband Luttinger--Kohn--Foreman formalism 
~\cite{Quantum_Theory_of_Cyclotron,For93} within the quantum technology computer-aided design (QTCAD) package~\cite{Interpretation,Analysis_and3D_TCAD}. The resulting confined states are then used in EDSR calculations, interband optical-transition modeling, and Lindblad master-equation simulations of the open-system dynamics~\cite{A_short_introduction}. Optical excitation is incorporated through a spectral-overlap formalism that accounts for the finite linewidths of both the external light source and the SAQD transitions.

Using a gated GaAs SAQD as a representative platform, we investigate how SAQD geometry, magnetic-field strength, transition wavelength, and illumination power affect electrically driven spin control and photon-induced Rabi-oscillation damping. The results reveal a competition between intrinsic optical-transition strength and spectral overlap with the illumination source, highlighting the need to co-optimize SAQD geometry, spin-control conditions, and the illumination spectrum.

The remainder of the paper is organized as follows. Sec.~II presents the theoretical and numerical modeling framework, including the electronic-structure model, spin-dynamics formalism, optical-transition analysis, and open-system dynamics. Sec.~III presents an application of the framework to the GaAs SAQDs, with emphasis on photon-induced damping of electrically driven hole-spin Rabi oscillations under different device and operating conditions. This section also discusses the implications for spin--photon quantum technologies and device optimization. Finally, Sec.~IV summarizes the main conclusions.

\section{Theoretical Framework}

\subsection{Device Platform and Modeling Objectives}

The objective of this work is to establish a predictive quantum-device simulation workflow for analyzing spin--photon interfaces based on semiconductor SAQDs. The framework is designed to quantitatively connect experimentally controllable device parameters, including SAQD geometry, gate voltages, magnetic field, optical wavelength, and optical power, to qubit performance metrics such as spin-control characteristics, optical-transition rates, and Rabi-oscillation decay times.

As a representative device, we consider the GaAs SAQD shown in Fig.~1(a). Its geometry is motivated by experimentally realized structures reported in Refs.~\cite{Excitonic_lifetimes,Cross-sectional,Highly_uniform,Single_photon_response_in_GaAs_quantum}. The GaAs island is surrounded by AlGaAs barriers in all three spatial dimensions, providing confinement for both electrons and holes. The six metallic gates deposited above the heterostructure [Fig.~1(b)] enable electrostatic tuning of the confinement potential and the implementation of EDSR. The modeling methodology developed here is general and applicable to arbitrary SAQD designs.

The simultaneous confinement of electrons and holes makes SAQDs particularly attractive for spin--photon quantum technologies. Spin qubits can be implemented using either confined electron states or confined hole states. In this work, we focus on hole-spin qubits because of their strong spin-orbit coupling and favorable EDSR characteristics.

A static magnetic field
\begin{equation}
\mathbf{B}=(B_x,B_y,B_z)=(0,0,B_0)
\label{eq1}
\end{equation}
is applied to define the spin quantization axis, as illustrated in Fig.~1(c). Coherent spin manipulation is achieved through EDSR~\cite{Electrically_driven,Electric-dipole-induced,Electrical_operation}. Under an oscillating gate voltage, the confined hole undergoes periodic orbital motion. In the presence of spin-orbit coupling and broken time-reversal symmetry by $\textbf{B}$, this orbital motion generates an effective oscillating magnetic field that drives coherent spin rotations on the Bloch sphere [Fig.~1(c)].

To answer the central question in this work, we develop an integrated workflow that combines (i) electronic-structure calculations,
(ii) spin-qubit and EDSR simulations, (iii) optical-transition modeling and (iv) open-quantum-system dynamics. The workflow is schematically shown in Fig.~2. It consists of a spin-qubit simulator and a spin–photon simulator that communicate through electronic eigenstates, optical matrix elements, and transition rates. The resulting methodology enables device-informed estimates of experimentally measurable quantities including Rabi frequencies, optical-transition rates, and photon-induced Rabi-oscillation decay times.
\begin{figure}
\includegraphics[width=0.50\textwidth]{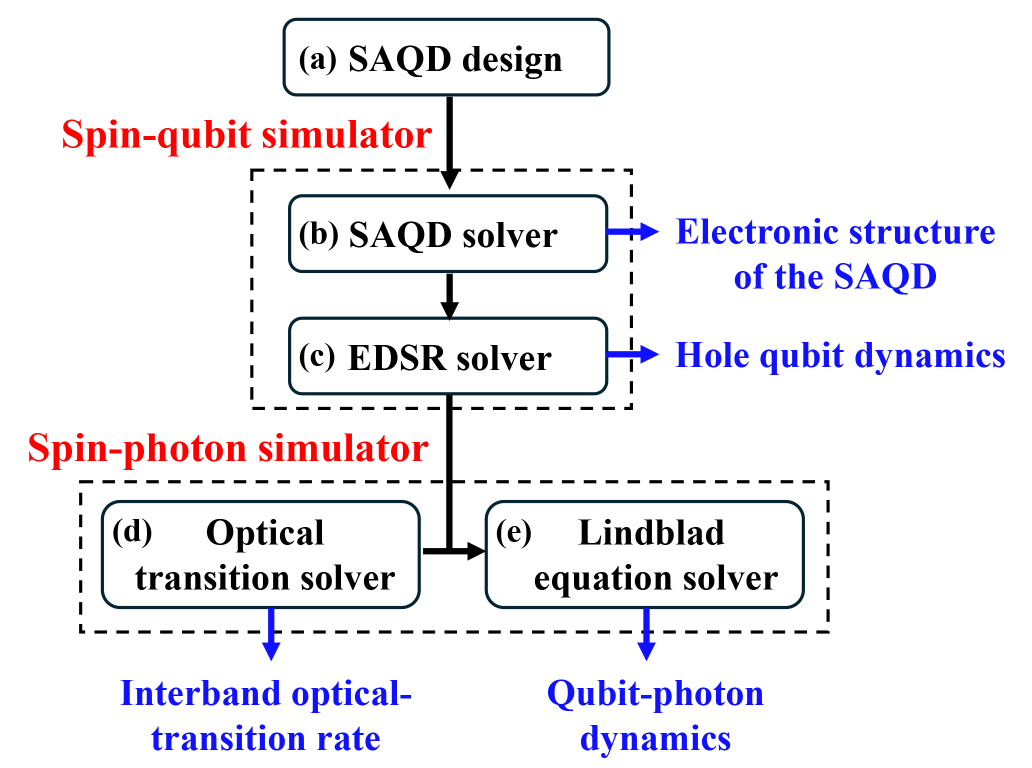}
\caption{\label{fig:Figure_2} The modeling workflow consists of a spin-qubit simulator and a spin–photon simulator. In the spin-qubit simulator, QTCAD is used to (a) define the SAQD structure and device geometry, (b) solve the single-particle Schrödinger equation using the finite-element method, and (c) simulate EDSR-driven spin dynamics. In the spin--photon simulator, (d) interband optical-transition rates are calculated using Fermi’s golden rule, and (e) the optically perturbed spin dynamics are determined by solving the Lindblad master equation. Outputs from each step are shown in blue.}
\end{figure}

\subsection{Electronic Structure of the SAQD}

The predictive capability of the framework begins with an accurate determination of the confined electronic states of the SAQD. These states are obtained using the Quantum Technology Computer-Aided Design package (QTCAD$^{\circledR}$)~\cite{Analysis_and3D_TCAD,Robust,Interpretation}, which self-consistently solves the Poisson equation and single-particle Schrödinger equation for arbitrary three-dimensional semiconductor nanostructures using the finite-element method (FEM). For the  GaAs/Al$_{0.3}$Ga$_{0.7}$As SAQD system considered in the next section, the electrostatic confinement is determined by both the heterostructure of the material and the externally applied gate voltages.

Within the envelope-function approximation~\cite{spin_orbit_coupling,Coherent_Optical}, the full electron and hole wave functions are expressed as
\begin{eqnarray}
&&\Psi_{n} (\textbf{r}) = \langle n\lvert\textbf{r}\rangle = \sum\limits_{m_J} {{F_{n,{m_J}}}(\textbf{r}){u_{J,{m_J}}}(\textbf{r})} \label{eq2}, \\
&&\Psi_{n'} (\textbf{r}) = \langle n'\lvert\textbf{r}\rangle  = \sum\limits_{m_{J}} {{F_{n',{m_{J}}}}(\textbf{r}){u_{J,{m_{J}}}}(\textbf{r})},\label{eq3}
\end{eqnarray}
where $n$ and $n'$ label the confined electron and hole states, respectively. The total angular-momentum quantum number in GaAs is $J=1/2$ for the conduction band (CB) and $J=3/2$ for the valence band (VB). $F_{n,m_J}$ and $F_{n',m_J}$ denote the slowly varying envelope functions, while $u_{J,m_J}$ are the cell-periodic Bloch functions of the semiconductor. The envelope functions capture the effects of confinement and device geometry, whereas the Bloch functions encode the atomic-scale crystal symmetry. After elimination of the cell-periodic contribution~\cite{spin_orbit_coupling}, the electronic structure reduces to the effective-mass equation for electrons and a multiband Luttinger--Kohn--Foreman description for holes.

Confined electron states are obtained from the effective-mass Schrödinger equation
\begin{equation}
\sum_{m_J'}(\hat{H}_e)_{m_J,m_J'}
F_{n,m_J'}(\mathbf{r})
=
E_n
F_{n,m_J}(\mathbf{r}),
\label{eq4}
\end{equation}
where the matrix elements of $\hat{H}_e$ are given by
\begin{equation}
\begin{aligned}
(\hat{H}_e)_{m_J,m_J'}
=
&\Big[
(
U_e(\mathbf{r})
+
\frac{\hbar^2}{2}\mathbf{k}\cdot\mathbf{M}_e^{-1}\cdot\mathbf{k}
)
\delta_{m_J,m_J'}\\
&+
(\mathcal{H}_{Z,e})_{m_J,m_J'}
\Big],
\end{aligned}
\label{eq5}
\end{equation}
where $U_e(\mathbf{r})$ is the electrostatically modified confinement potential; $\textbf{k}=-i\nabla+ \frac{e}{\hbar} \mathbf{A}$ is the wave vector operator; $\mathbf{M}_e$ is the electron effective-mass tensor; $\hbar$ is the reduced Planck constant. We neglect strain effects, so the effective mass tensor is isotropic, with $M^{*}$=0.067 $m_0$ for GaAs, where $m_0$ is the bare electron mass. $\mathcal{H}_{Z,e}$ is the electron Zeeman term,
\begin{equation}
H_{Z,e}
=
\frac{\mu_B g^*}{\hbar} 
\mathbf{S}\cdot\mathbf{B},
\label{eq6}
\end{equation}
where $\mu_B$ is the Bohr magneton, $g^*=-0.44$ is the effective Landé $g$ factor for GaAs, and the spin operator is $\mathbf{S}=\frac{\hbar}{2}{\boldsymbol{\sigma}}$, where $\boldsymbol{\sigma}$ is the vector of Pauli matrices. The external magnetic field lifts the Kramers degeneracy and splits each confined level into two spin states, thereby defining the spin-qubit basis.

Because valence-band states exhibit strong heavy-hole/light-hole mixing, a multiband treatment is required. Hole states are therefore described using the four-band Luttinger--Kohn--Foreman Hamiltonian, and the corresponding effective equation is
\begin{equation}
\sum_{m_J'}(\hat{H}_h)_{m_J,m_J'}
F_{n',m_J'}(\mathbf{r})
=
E_{n'}
F_{n',m_J}(\mathbf{r}),
\label{eq7}
\end{equation}
where the matrix elements of $\hat{H}_h$ are given by
\begin{equation}
\begin{aligned}
(\hat{H}_h)_{m_J,m_J'}
=
\Big[&
U_h(\mathbf{r})
\delta_{m_J,m_J'}
+
\sum_{\alpha,\beta}
k_\alpha
D^{\alpha,\beta}_{m_J,m_J'}
k_\beta\\
&+
(\mathcal{H}_{Z,h})_{m_J,m_J'}
\Big].
\end{aligned}
\label{eq8}
\end{equation}
$U_{\text{h}}$ is the hole confinement potential and D denotes the D-matrix in the Luttinger--Kohn--Foreman Hamiltonian, summarized in Appendix~\ref{appendix:A}. The Zeeman term for confined holes is
\begin{equation}
H_{Z,h}
=
-2\frac{\mu_B\kappa}{\hbar}
\mathbf{J}_{3/2}\cdot\mathbf{B} - 2\frac{\mu_B q}{\hbar}
\mathcal{J}\cdot\mathbf{B},
\label{eq9}
\end{equation}
where $\kappa=1.2$ and $q=0.01$ are the hole Zeeman parameters for GaAs, $\mathbf{J}_{3/2}$ denotes the spin-$3/2$ angular-momentum matrices, and $\mathcal{J}=(J_x^3,J_y^3,J_z^3)$. For the magnetic field $\mathbf{B}$ of Eq.~(\ref{eq1}), only the z-component of the angular-momentum operator contributes. The resulting eigenstates naturally incorporate heavy-hole/light-hole mixing, magnetic-field effects, and realistic device confinement. These states [Eqs.~(\ref{eq2}) and (\ref{eq3})] provide the foundation for the spin-dynamics and optical-transition calculations described in the following subsections.

\subsection{Spin-Control: Electric-Dipole Spin Resonance}

The confined states obtained from the electronic-structure calculations [Eqs.~(\ref{eq2}) and (\ref{eq3})] provide the foundation for implementing spin qubits and evaluating their coherent dynamics. In the present work, the qubit is encoded in the two lowest-energy hole states, $\lvert0'\rangle$ and $\lvert1'\rangle$, which result from the Zeeman splitting of the hole ground state.

Coherent manipulation of the hole-spin qubit is achieved through EDSR. Compared with conventional spin resonance techniques, EDSR enables all-electrical spin control using gate voltages and is therefore attractive for scalable semiconductor quantum hardware. In GaAs SAQDs, the strong spin-orbit coupling and heavy-hole/light-hole mixing convert periodic orbital hole motion into an effective oscillating magnetic field as seen by the hole, leading to spin rotations [see Fig.~1(c)].

To realize EDSR, an external time-dependent voltage is applied to gate G6 [see Fig.~1(b)],
\begin{equation}
V_{G6}(t)=V_0+\delta V \cos(\omega t),
\label{eq10}
\end{equation}
where $V_0$ denotes the operating point and $\delta V$ is the driving amplitude of the ac voltage. The resulting perturbation modifies the electrostatic confinement potential and generates a time-dependent qubit Hamiltonian
\begin{equation}
\hat{H}_{qb}(\mathbf{r},t)
=
\hat{H}_{eh}(\mathbf{r})
+
\delta \hat{U}(\mathbf{r,t}),
\label{eq11}
\end{equation}
where $\hat{H}_{eh}$ is the static electron--hole Hamiltonian constructed from the confined states obtained in the previous subsection. The time-dependent  perturbation inside the SAQD induced by $\delta V$ is denoted by $\delta \hat{U}(\mathbf{r,t})$.

To enable efficient numerical simulations, the Hamiltonian is projected onto a truncated basis consisting of the four lowest hole states and the two lowest electron states. In EDSR, the driving frequency $\omega$ is tuned to the hole-spin resonance (level spacing between $\lvert0'\rangle$ and $\lvert1'\rangle$), inducing Rabi oscillations of the hole spin. The same driving frequency is far detuned from the electron-spin resonance and therefore does not induce appreciable Rabi oscillations in electron states. Nevertheless, both the electron and hole states are retained in the truncated basis to describe optical transitions between them. Within the truncated basis, the gate-induced perturbation is obtained from the difference between electrostatic solutions corresponding to the instantaneous gate voltage and the reference operating point,
\begin{equation}
\delta U_{m'n'}(t)
=
U_{m'n'}
\left(
V_0+\delta V\cos\omega t
\right)
-
U_{m'n'}(V_0),
\label{eq12}
\end{equation}
where the matrix elements are evaluated as
\begin{equation}
U_{m'n'}(V_{G6})
=
\int d^3r \,
\Psi_{m'}^*(\mathbf r)
U_h(\mathbf r,V_{G6})
\Psi_{n'}(\mathbf r).
\label{eq13}
\end{equation}
The electrostatic potential entering Eq.~(\ref{eq13}) is determined self-consistently from the Poisson equation under the corresponding gate-voltage boundary conditions. This procedure establishes a direct connection between experimentally applied control voltages and the effective qubit Hamiltonian Eq.~(\ref{eq11}).

The coherent dynamics of the electrically driven qubit is governed by the von Neumann equation~\cite{Open_quantum_systems},
\begin{equation}
\frac{d\hat{\rho}_0}{dt}
=
-\frac{i}{\hbar}
\left[
\hat{H}_{qb},
\hat{\rho}_0
\right],
\label{eq14}
\end{equation}
where $\hat{\rho}_0$ denotes the density matrix of the isolated device.
The occupation probability of each state is obtained from the diagonal element of the density matrix,
\begin{equation}
O_j(t)
= \langle j\lvert \hat{\rho}_0(t)
\lvert j \rangle ,
\label{eq15}
\end{equation}
where
$\{\lvert j\rangle\}=
\{\lvert0'\rangle,\lvert1'\rangle,\lvert2'\rangle,\lvert3'\rangle,\lvert0\rangle,\lvert1\rangle\}
$
is the truncated basis.
Solving Eqs.~(\ref{eq10})--(\ref{eq14}) yields the Rabi oscillations and coherent spin-control characteristics that form the baseline performance metrics of the qubit prior to coupling with optical fields.

\subsection{Optical Coupling}

Optical interband transitions between the QD-confined hole and electron states are essential for spin initialization, readout, and spin--photon interfacing. At the same time, optical excitation introduces additional dissipative channels that degrade qubit coherence. Quantitative evaluation of these competing effects requires a microscopic description of interband optical transitions.

Within the present framework, optical-transition rates are evaluated using Fermi's golden rule~\cite{Quantum_Mechanics}. The spontaneous-emission rate associated with a transition between the confined electron and hole states, $\lvert n\rangle$ and $\lvert n'\rangle$, is
\begin{equation}
\Gamma_{n'n}^{0}
=
\int
\gamma_{n'n}^{0}
\,d^{3}N_{\mathrm{ph}},
\label{eq16}
\end{equation}
where $d^{3}N_{\mathrm{ph}}$ is the differential measure over
photon modes, including the photon frequency, propagation direction, and polarization. The mode-resolved transition rate $\gamma_{n'n}^{0}$ is given by
\begin{equation}
\gamma_{n'n}^{0}
=
\frac{2\pi}{\hbar}
\left \lvert
\left\langle
\Psi_{n'}
\left \lvert
\hat H_{\mathrm{int}}
\right \lvert
\Psi_n
\right\rangle
\right \lvert^{2}
\delta
\left[
(E_n-E_{n'})
-\hbar\omega
\right],
\label{eq17}
\end{equation}
where $\hat H_{\mathrm{int}}$ denotes the effective particle-photon interaction operator for a given photon mode. The calculation of the interband optical matrix elements $\left \lvert \left\langle {{\Psi _{n'}}} \right \lvert \hat H_{\mathrm{int}}\left \lvert {{\Psi_{n}}} \right\rangle  \right \lvert ^2$ is included in Appendix~\ref{Appendix:B}. They naturally incorporate the effects of quantum confinement, heavy-hole/light-hole mixing, and device geometry. 

In the rest of the work, for simplicity, we employ an independent-particle approximation to calculate the interband transitions. In Sec.~III, we investigate a SAQD whose height is smaller than the characteristic exciton Bohr radius, indicating strong confinement along the growth direction~\cite{StobbePRB2009}; the larger lateral dimensions of the SAQD are in the intermediate-to-strong confinement regime~\cite{Very_large,Quantum_Theory_of_the_Optical}. In the strongly confined regime, one may neglect the electron--hole exciton binding energy as a first approximation, since quantum confinement dominates the energy spectrum. Nevertheless, for the SAQD we consider, this approximation may lead to a small overestimation of the optical-transition energy. The calculated transition wavelengths and spectral-overlap-dependent Rabi-oscillation decay times should be interpreted accordingly. 

The resulting transition rates provide the fundamental link between the microscopic electronic structure and the experimentally measurable optical response. Because the optical matrix elements are computed from realistic confined states, the framework enables the predictive evaluation of optical coupling across a wide range of device geometries and operating conditions.

\begin{figure*}
\includegraphics[width=1
\textwidth]{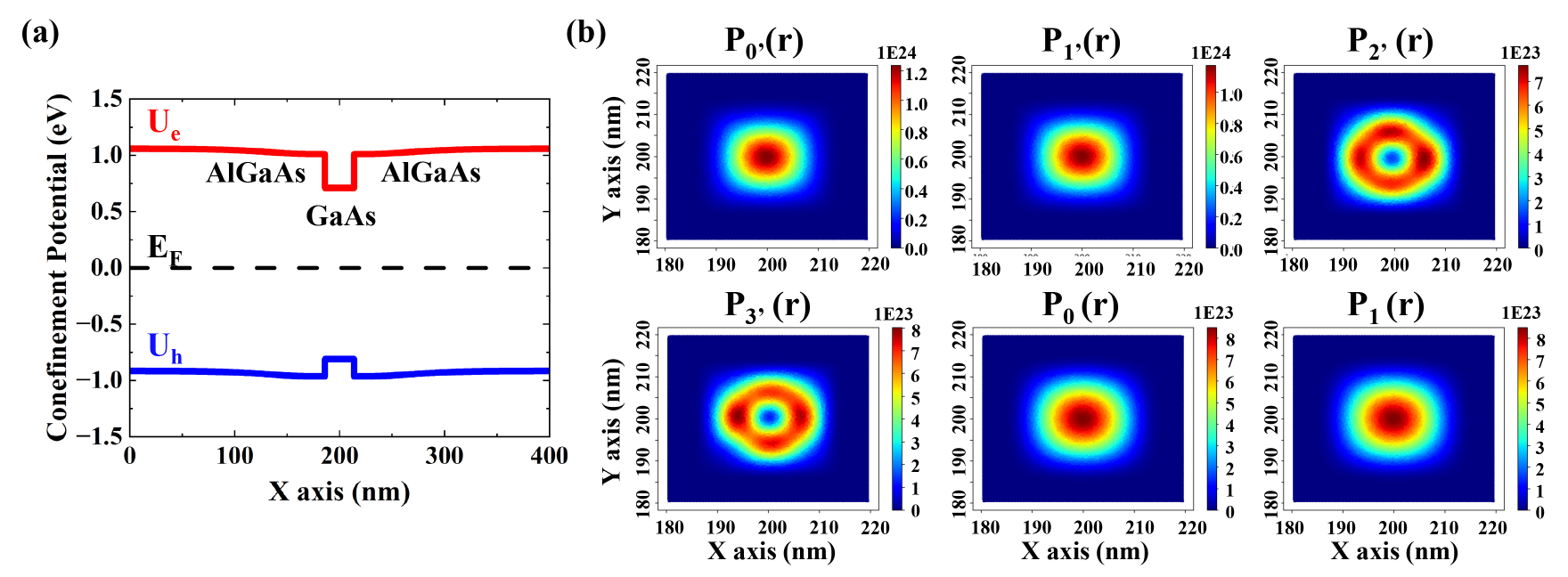}
\caption{\label{fig:Figure_3} Results of the SAQD solver [box (b) of Fig.~2]. For this system [Figs.~1(a) and 1(b)], the device parameters are listed in Table \ref{tab:table1}, a gate voltage of $-140$ mV is applied to the top gates G1--G6, and $B_0$ is set to $2~\mathrm{T}$. (a) The calculated electron and hole confinement potentials,
$U_e$ and $U_h$ together with the Fermi level $E_F$, along the $x$ axis through the center of the Al$_{0.3}$Ga$_{0.7}$As/GaAs/Al$_{0.3}$Ga$_{0.7}$As device structure. (b) The calculated probability densities of the four lowest-energy hole states 
$P_{n'}(\mathbf r)=\lvert \Psi_{n'}(\mathbf r) \lvert^2$ ($n'=0',1',2',3'$) and the two lowest-energy electron states $P_{n}(\mathbf r)=\lvert\Psi_{n}(\mathbf r)\lvert^2$ ($n=0,1$) with units of $m^{-3}$. The calculated eigenenergies of these states are listed in Table \ref{tab:table2}.}
\end{figure*}

\subsection{Open-System Dynamics and Optical Decoherence}

To investigate the influence of optical excitation on qubit performance, the spin qubit is treated as an open quantum system interacting with an external electromagnetic environment.
The optical field couples the confined hole and electron states through absorption and emission processes. The broadband illumination is treated as an incoherent Markovian reservoir with independent absorption and emission channels in a homogeneous GaAs photonic environment. After integrating out the photonic degrees of freedom, the reduced density matrix of the quantum dot obeys the Lindblad master equation~\cite{A_short_introduction},
\begin{equation}
\begin{aligned}
\frac{d\hat{\rho}}{dt}
=
&-\frac{i}{\hbar}
\left[
\hat{H}_{qb},
\hat{\rho}
\right]
\\
&
+
\sum_{n,n'}
\Bigg[
-\frac12
\left\{
\left(\hat L_{nn'}^{\mathrm{ab}}\right)^{\dagger}
\hat L_{nn'}^{\mathrm{ab}},
\hat\rho
\right\}
+
\hat L_{nn'}^{\mathrm{ab}}
\hat\rho
\left(\hat L_{nn'}^{\mathrm{ab}}\right)^{\dagger}
\\
&
-\frac12
\left\{
\left(\hat L_{n'n}^{\mathrm{em}}\right)^{\dagger}
\hat L_{n'n}^{\mathrm{em}},
\hat\rho
\right\}
+
\hat L_{n'n}^{\mathrm{em}}
\hat\rho
\left(\hat L_{n'n}^{\mathrm{em}}\right)^{\dagger}
\Bigg].
\end{aligned}
\label{eq18}
\end{equation}
The summation runs over all allowed optical transitions between the confined electron states $\lvert n\rangle$ and hole states $\lvert n'\rangle$. The jump operators $\hat{L}^{\mathrm{ab}}_{n n'}$ and $\hat{L}^{\mathrm{em}}_{n' n}$ describe photon absorption and emission processes, respectively, and are constructed from the absorption and emission rates $\Gamma_{nn'}$ and $\Gamma_{n'n}$ (see below). The jump operators are defined as
\begin{align}
\hat{L}^{\mathrm{ab}}_{n n'}
&=
\sqrt{\Gamma_{n n'}}\, \lvert n\rangle\langle n'\lvert ,
\\
\hat{L}^{\mathrm{em}}_{n' n}
&=
\sqrt{\Gamma_{n' n}}\, \lvert n'\rangle\langle n \lvert .
\end{align}
The Lindblad operators couple the confined hole-state and electron-state subspaces and therefore provide a microscopic mechanism through which optical excitation modifies the spin dynamics of the qubit. In the absence of optical coupling, the jump operators vanish and Eq.~(\ref{eq18}) reduces to the closed-system dynamics of Eq.~(\ref{eq14}).

The open-system formulation establishes a direct connection between optical-transition physics and experimentally relevant qubit metrics, including population dynamics and Rabi-oscillation decay times. Most importantly, it enables a quantitative evaluation of how device geometry, magnetic field, optical wavelength, and optical power
influence photon-induced Rabi-oscillation damping.

\subsection{Integrated Workflow}

The complete simulation workflow indicated in Fig.~2 combines the electronic-structure, spin-control, optical-coupling, and open-system dynamics within a unified computational framework. Information flows sequentially through the framework via confined eigenstates, electrostatic potentials, optical matrix elements, and transition rates. For a specified device geometry and operating condition, the framework predicts confined electron and hole states; qubit energies and EDSR response; optical-transition strengths; photon-induced excitation and relaxation rates; and Rabi-oscillation decay times.

These quantities constitute the key design metrics for spin--photon quantum hardware. In the following section, the methodology is applied to realistic GaAs SAQDs to establish a design for minimizing photon-induced Rabi-oscillation damping while maintaining efficient spin control and optical accessibility.

\section{Device-Level Design Studies and Results}

In this section, the computational framework presented in Sec.~II is applied to a realistic GaAs SAQD device. The objective is to establish quantitative relationships among device geometry, magnetic-field strength, optical operating conditions, and spin-qubit performance. Particular attention is devoted to identifying design parameters that maximize spin controllability while minimizing photon-induced Rabi-oscillation damping.

\subsection{Qubit-State Engineering}
{\setlength{\textfloatsep}{5pt}
\begin{table}[t]
\caption{\label{tab:table1}
Device and control parameters for the system in Fig.~1(a). }
\begin{ruledtabular}
\renewcommand{\arraystretch}{1.2}
\begin{tabular}{lcc}
\textrm{Parameter}  & \textrm{Value} \\
\colrule
Top area of SAQD  & 1253$~\mathrm{nm}$$^2$ \\
Bottom area of SAQD & 400$~\mathrm{nm}$$^2$ \\
QD height ($h_{QD}$) & 7.6$~\mathrm{nm}$ \\
Gate voltages $V_{G1}$--$V_{G5}$  & -140 mV \\
Gate voltage operating range $V_{G6}$ & -$141$ to -$139~\text{mV}$ \\
Magnetic-field strength ($B_0$) &  $1~\mathrm{T}$ or $2~\mathrm{T}$\\
EDSR resonance frequency ($f_{Res}$) &~6.2$~\mathrm{GHz}$ ($B_0=1~\mathrm{T}$)\\
  & 14.2$~\mathrm{GHz}$ ($B_0=2~\mathrm{T}$) \\
\end{tabular}
\end{ruledtabular}
\end{table}
}
\begin{figure*}
\includegraphics[width=1.00\textwidth]{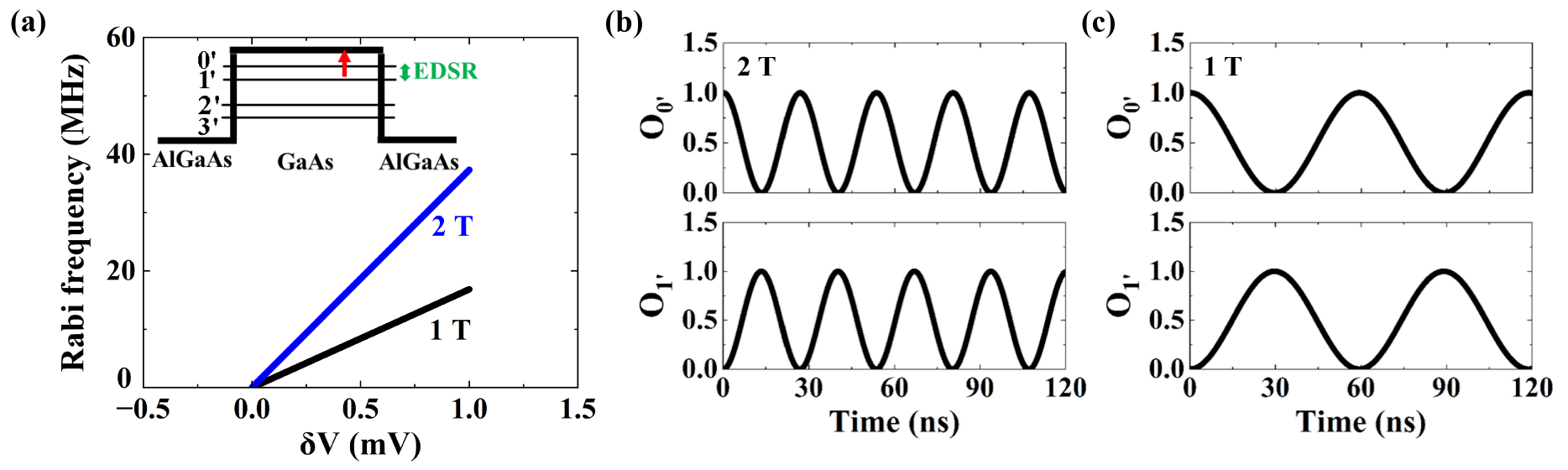}
\caption{\label{fig:Figure_4} Results of the EDSR solver [box (c) in Fig.~2]. For this system [Figs.~1(a) and 1(b)], the device parameters are listed in Table~\ref{tab:table1}. A dc bias of $-140~\mathrm{mV}$ is applied to all the top gates G1--G6 and an additional ac voltage $\delta V\cos(\omega t)$ is applied to gate G6 to drive EDSR with $0\leq\delta V\leq1~\mathrm{mV}$. (a) Hole-qubit Rabi frequency versus the ac driving amplitude $\delta V$, with $B_0=1~\mathrm{T}$ (black) and $B_0=2~\mathrm{T}$ (blue). The inset in panel (a) schematically shows the lowest-energy hole states where the two-level system for the spin qubit is indicated by EDSR (green). (b,c) Rabi oscillations of the occupation probabilities at $B_0=2~\mathrm{T}$ and $B_0=1~\mathrm{T}$, respectively.}
\end{figure*}

For the SAQD shown in Fig.~1, the device and simulation parameters, including the SAQD dimensions, the applied gate voltages, the magnetic-field strength, and the EDSR resonance frequency, are summarized in Table~\ref{tab:table1}. Motivated by the experimental SAQD reported in Refs.~\cite{Excitonic_lifetimes, Highly_uniform}, we adopt a similar SAQD geometry with the same top area and the same QD height, and model the etched SAQD as an inverted truncated pyramid. Fig.~3(a) shows the calculated electron and hole effective confinement potentials $U_e$ and $U_h$ along the x-direction across the middle of the SAQD, when a voltage $V_0=-140~\mathrm{mV}$ is applied to gates G1--G6. This $V_0$ is set so that both the hole and the electron are simultaneously confined in the center of the SAQD. Fig.~3(b) shows the calculated probability densities $P_{n}(\mathbf r)=\lvert \Psi_{n}(\mathbf r)\lvert ^2=\sum\limits_{m_J} \lvert F_{n,m_J}(\mathbf r) \lvert^2$  and $P_{n'}(\mathbf r)=\lvert \Psi_{n'}(\mathbf r)\lvert^2=\sum\limits_{m_J}\lvert F_{n',m_{J}}(\mathbf r)\lvert^2$, for the two lowest-energy electron states (labeled $n=0,1$) and the four lowest-energy hole states (labeled $n'=0',\cdots 3'$), in the central cross-sectional plane of the device [see Fig.~1(a)]. The calculated eigenenergies of these states are listed in Table \ref{tab:table2}. The dark blue region corresponds to the Al$_{0.3}$Ga$_{0.7}$As barrier. The ground states $\Psi_{0'}$ and $\Psi_{1'}$, as well as $\Psi_0$ and $\Psi_1$, are $s$-like, whereas the excited states $\Psi_{2'}$ and $\Psi_{3'}$ are $p$-like and exhibit a local minimum at the center. In the absence of magnetic field, the two electron states ($n=0,1$) are spin degenerate, and the same applies to the hole-state pairs $n'=0',1'$ and $n'=2',3'$.

The calculated wave functions exhibit strong confinement within the GaAs island and significant heavy-hole/light-hole mixing arising from the multiband valence-band structure. This mixing plays an important role in determining both the EDSR response and the optical-transition strengths. The resulting energy spectrum provides the foundation for all subsequent spin--control and spin--photon simulations. The confined electron states serve as intermediate states for optical excitation and emission processes. The energy differences between the confined hole and electron states determine the optical-transition frequencies and therefore govern the spectral overlap with external optical sources.

Finally, the spin qubit is implemented in the two lowest-energy hole states denoted by $\lvert 0'\rangle$ and $\lvert 1'\rangle$, indicated by the red upward-pointing arrow in the inset of Fig.~4(a).
 
{\setlength{\textfloatsep}{5pt}
\begin{table}[t]
\caption{\label{tab:table2}
Eigenenergies of the confined hole and electron states of the SAQD. The system parameters are listed in Table \ref{tab:table1}. Note that, in our convention, the hole-state energies are negative because they
lie below the material Fermi level. The listed eigenenergies are rounded.}
\begin{ruledtabular}
\renewcommand{\arraystretch}{1.2}
\begin{tabular}{lcc}
\textrm{Eigenenergy}  & \textrm{Value ($B_0=1~\mathrm{T}$)} & \textrm{Value ($B_0=2~\mathrm{T}$)}\\
\colrule
$E_{0'}$  &  -0.82754$~\mathrm{eV}$ &  -0.82812$~\mathrm{eV}$\\
$E_{1'}$  &  -0.82756$~\mathrm{eV}$ &  -0.82818$~\mathrm{eV}$\\
$E_{2'}$  &  -0.83176$~\mathrm{eV}$ &  -0.83220$~\mathrm{eV}$\\
$E_{3'}$  &  -0.83228$~\mathrm{eV}$ &  -0.83324$~\mathrm{eV}$\\
$E_{0}$   &   0.77194$~\mathrm{eV}$ &   0.77326$~\mathrm{eV}$\\
$E_{1}$   &   0.77196$~\mathrm{eV}$ &   0.77331$~\mathrm{eV}$\\
\end{tabular}
\end{ruledtabular}
\end{table}
}
\begin{figure*}
\includegraphics[width=1\textwidth]{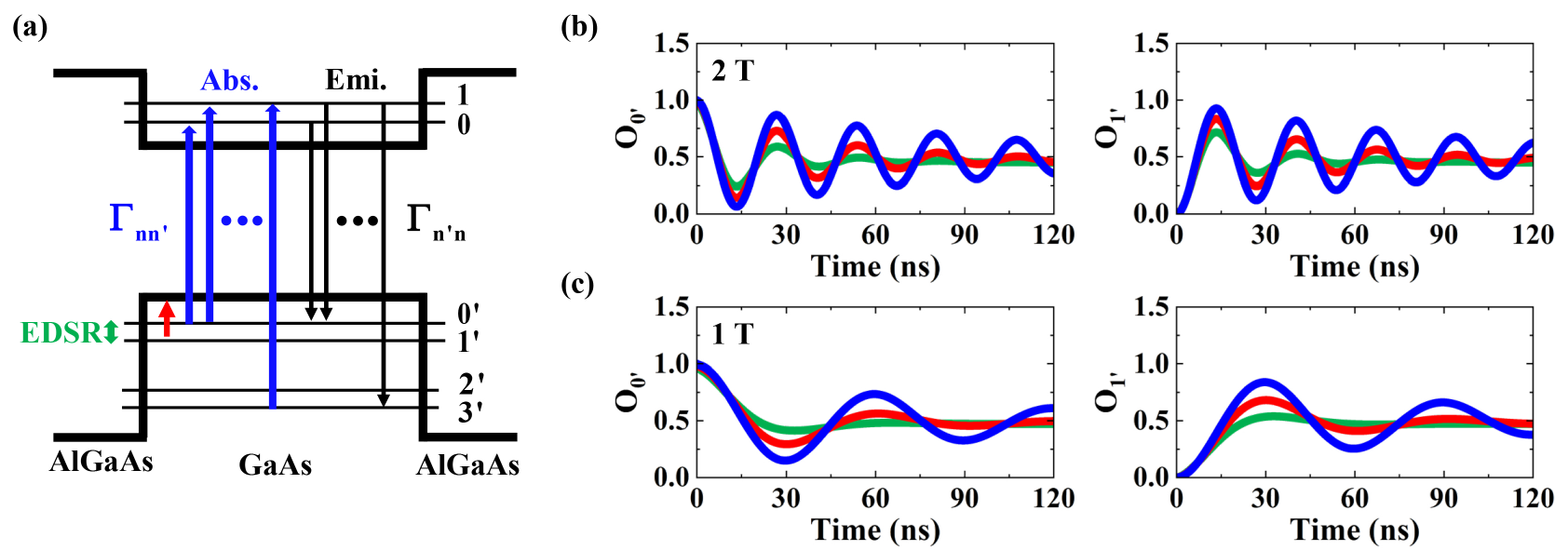}
\caption{\label{fig:Figure_5} (a) Schematic showing the optical transitions (upward and downward arrows). The qubit is initialized in $\lvert 0'\rangle$ and manipulated via EDSR while the SAQD simultaneously interacts with external electromagnetic radiation near the interband transition energy. The calculated qubit-state occupation probabilities for (b) $B_0$ = 2$~\mathrm{T}$ and (c) $B_0$ = 1$~\mathrm{T}$ under LED powers of 0.3, 0.75, and 1.5$~\mathrm{mW}$, shown in blue, red, and green, respectively.}
\end{figure*}

\subsection{Electric-Dipole Spin Resonance}

The coherent control characteristics of the hole-spin qubit are evaluated using the EDSR module described in Sec.~II. According to Eq.~(\ref{eq10}), EDSR is driven by a time-dependent gate voltage applied to gate G6 [Fig.~1(b)]. The calculated time evolution of the qubit is shown in Fig.~4. Experimentally, the amplitude of the time-dependent perturbing voltage $\delta V$ [Eq.~(\ref{eq10})] is carefully controlled to prevent thermal effects that can induce spin-qubit decoherence~\cite{Operation_of, Universal_quantum}. Based on the device dimensions and drive amplitudes reported in Ref.~\cite{A_germanium_hole}, the corresponding electric-field scale is of order $10^4~\mathrm{V/m}$. The EDSR driving frequency is determined by the hole-spin resonance frequency $f_{Res}=\lvert E_{1'}-E_{0'}\lvert / h$. 

The simulations demonstrate coherent Rabi oscillations [Figs.~4(b) and 4(c)] between the two qubit states $\lvert 0'\rangle$ and $\lvert 1'\rangle$, confirming that the time-dependent gate-induced electric field efficiently drives spin rotations through the spin-orbit coupling~\cite{Electrical_operation} that is included in the Luttinger--Kohn--Foreman Hamiltonian as the off-diagonal terms in Eq.~(\ref{eqA1}). Fig.~4(a) plots the calculated Rabi frequency as a function of the ac driving amplitude $\delta V$, revealing an approximately linear dependence within the investigated driving range, consistent with the experimental results reported for a GaAs hole-spin qubit in Ref.~\cite{Coherence_Characteristics}.

The driven-qubit dynamics are obtained by numerically solving the von Neumann equation for the density matrix in Eq.~(\ref{eq14}) using the QuTiP package~\cite{qutip_1,qutip_2}. Figs.~4(b) and 4(c) show the calculated occupation probabilities of the qubit states. The hole spin is initialized in the $\lvert 0'\rangle$ state, and the driving angular frequency $\omega$ is set equal to the resonance angular frequency $\omega_{Res}=2\pi f_{Res}$ of the $\lvert 0'\rangle \leftrightarrow \lvert 1'\rangle$ transition. EDSR induces periodic rotations of the spin within the qubit subspace, as reflected in the time-dependent occupation probabilities $O_{0'}$ and $O_{1'}$. Since the transitions to the higher excited states are off resonance, the occupation probabilities of the leakage states, $O_{2'}$ and $O_{3'}$, are negligible and therefore not shown in Fig.~4. The calculated hole-spin Rabi frequencies are approximately $37.3~\mathrm{MHz}$ at $B_0=2~\mathrm{T}$ and $16.8~\mathrm{MHz}$ at $B_0=1~\mathrm{T}$. These values are comparable to the experimentally reported Rabi-frequency range of 10--40$~\mathrm{MHz}$ for a gate-driven GaAs heavy-hole spin qubit at 0.896$~\mathrm{T}$~\cite{Coherence_Characteristics}. These results establish that GaAs SAQDs provide an attractive platform for electrically controlled spin qubits with relatively fast manipulation rates.

\subsection{Optical Coupling and Photon-Induced Rabi-Oscillation Damping}

Having established coherent spin control, we next investigate how spontaneous emission and external optical illumination affect qubit dynamics. Spontaneous emission couples the confined electron and hole states, while external illumination induces optical interband transitions, thereby creating a coupling channel between the spin qubit and the electromagnetic environment. Here we investigate lifetime-limited SAQD transitions and neglect additional broadening mechanisms such as spectral diffusion~\cite{QD_Spectral_Diffusion}, phonon-induced broadening~\cite{QD_Phonon_Broadening}, and inhomogeneous broadening~\cite{QD_Inhomogeneous_Broadening}.

In the absence of LED illumination, the SAQD spontaneous-emission rates defined in Eq.~(\ref{eq16}) are calculated and summarized in Tables~\ref{tab:table3} and \ref{tab:table4}. The photon wavelengths are determined by the energy difference ($E_n - E_{n'}$) between the confined electron and hole states taken from Table~\ref{tab:table2}. If an electron is initially prepared in the $\lvert 0\rangle$ state, it can relax to the $\lvert 0'\rangle$ and $\lvert 1'\rangle$ states through spontaneous emission. The corresponding spontaneous-emission lifetime in Table~\ref{tab:table3} is estimated as $\left[\Gamma^0_{sp}\right]^{-1}=\left[{\Gamma^0_{0'0}+\Gamma^0_{1'0}}\right]^{-1}=458.3~\text{ps}$. This value is comparable to the electron--hole recombination lifetime of $390~\text{ps}$ measured for a SAQD of similar size~\cite{Excitonic_lifetimes}. Given the differences between the theoretical model and the experimental conditions such as the absence of an applied gate voltage in the experimental SAQD, this level of consistency is acceptable.

In the optical-cycling process, optical excitation produces carrier transitions across the band gap, while spontaneous emission returns the carrier to the confined hole states. These repeated excitation and relaxation processes act as an effective noise source that degrades the coherence of the spin qubit. We consider a typical infrared LED source with a central wavelength $\lambda_{\mathrm{LED}}=790~\mathrm{nm}$ and a spectral width $\Delta\lambda_{\mathrm{LED}}=20~\mathrm{nm}$~\cite{SLED_EXS210067-03}. We assume an effective focused-beam area of $A_{\mathrm{beam}}=5\times10^{-12}~\mathrm{m}^2$. For comparison, the $1.3~\mu\mathrm{m}$ FWHM Gaussian spot reported in Ref.~\cite{Stark-shift} corresponds to an illuminated area of approximately $2\times10^{-12}~\mathrm{m}^2$. The optical power ranges from $0.3$ to $1.5~\mathrm{mW}$, corresponding to local optical intensities at the SAQD plane that range from $6\times 10^7$ to $3\times10^{8}~\mathrm{W\,m^{-2}}$. Here, the optical power refers to the power contained within the effective beam area. The normalized spectral power density is assumed to be a Gaussian profile, $D_{LED}(\nu)=\frac{1}{\sqrt{2 \pi} \sigma_{LED}} \text{exp}\left[-\frac{(\nu-\nu_{LED})^2}{2\sigma_{LED}^2} \right]$, where $\nu_{LED}=\frac{c}{\lambda_{\mathrm{LED}}}$ is the central spectral frequency, $\sigma_{LED}=\frac{\Delta \nu_{LED}}{2 \sqrt{2 ln2}}$ is the standard deviation determined by the full-width at half-maximum of the spectrum.

Accounting for the stochastic nature of the environmental light source interacting with the SAQD system, we consider the mean photon number $\langle N_{nn'}\rangle$, which describes the time-averaged occupation of the electromagnetic mode and is derived in Appendix~C. The absorption and emission rates~\cite{Quantum_Mechanics} are given by $\Gamma_{nn'} =\langle N_{nn'}\rangle \Gamma^0_{n'n}$ and $\Gamma_{n'n} = \left[\langle N_{nn'}\rangle + 1\right] \Gamma^0_{n'n}$, so the calculated absorption rates at $B_0=2~\text{T}$ and $1~\text{T}$ are listed in Tables~\ref{tab:table3} and~\ref{tab:table4}, respectively.

{\setlength{\textfloatsep}{5pt}
\begin{table}
\caption{\label{tab:table3}
Calculated optical-transition parameters between the $(n',n)$ states of the SAQD under a magnetic field of $B_0=2~\mathrm{T}$, including wavelength ($\lambda_{\mathrm{QD}}$), spontaneous-emission rates ($\Gamma^0_{n'n}$), and absorption rates ($\Gamma_{nn'}$) at an LED power of $1.5~\mathrm{mW}$.}
\begin{ruledtabular}
{\renewcommand{\arraystretch}{1.2}
    \begin{tabular}{lccccc}
    (n',n) & $\lambda_{\mathrm{QD}}$ ($\mathrm{nm}$)  & $\Gamma^0_{n'n}$ ($\mathrm{Hz}$) & $\Gamma_{nn'}$ ($\mathrm{Hz}$)\\
    \hline
    $(0',0)$ & 775 & $2.07\times10^9$ & $8.18\times10^7$ \\ 
    $(0',1)$ & 775 & $1.14\times10^8$ & $4.49\times10^6$  \\ 
    $(1',0)$ & 775 & $1.13\times10^8$ & $4.44\times10^6$  \\ 
    $(1',1)$ & 775 & $2.12\times10^9$ & $8.29\times10^7$  \\ 
    $(2',0)$ & 773 & $1.15\times10^7$ & $2.83\times10^5$  \\ 
    $(2',1)$ & 773 & $3.74\times10^6$ & $9.10\times10^4$ \\ 
    $(3',0)$ & 772 & $8.85\times10^6$ & $1.90\times10^5$  \\ 
    $(3',1)$ & 772  & $1.03\times10^7$ & $2.20\times10^5$ \\ 
    \end{tabular}
    }
\end{ruledtabular}
\end{table}
}
{\setlength{\textfloatsep}{5pt}
\begin{table}
\caption{\label{tab:table4}
Calculated optical-transition parameters between the $(n',n)$ states of the SAQD under a magnetic field of $B_0=1~\mathrm{T}$, including wavelength ($\lambda_{\mathrm{QD}}$), spontaneous-emission rates ($\Gamma^0_{n'n}$), and absorption rates ($\Gamma_{nn'}$) at an LED power of $1.5~\mathrm{mW}$.}
\begin{ruledtabular}
{\renewcommand{\arraystretch}{1.2}
    \begin{tabular}{lccccc}
    $(n',n)$ & $\lambda_{\mathrm{QD}}$ ($\mathrm{nm}$)  & $\Gamma^0_{n'n}$ ($\mathrm{Hz}$)   & $\Gamma_{nn'}$ ($\mathrm{Hz}$)\\
    \hline
    $(0',0)$ & 776   & $2.12\times10^9$ & $1.03\times10^8$ \\ 
    $(0',1)$ & 776   & $7.33\times10^7$ & $3.55\times10^6$  \\ 
    $(1',0)$ & 776   & $7.31\times10^7$ & $3.53\times10^6$  \\ 
    $(1',1)$ & 776   & $2.15\times10^9$ & $1.04\times10^8$  \\ 
    $(2',0)$ & 774   & $1.62\times10^6$ & $4.91\times10^4$  \\ 
    $(2',1)$ & 774   & $6.06\times10^5$ & $1.83\times10^4$ \\ 
    $(3',0)$ & 773   & $1.10\times10^6$ & $3.12\times10^4$  \\ 
    $(3',1)$ & 773   & $1.57\times10^6$ & $4.45\times10^4$ \\ 
    \end{tabular}
    }
\end{ruledtabular}
\end{table}
}

The optically damped spin dynamics is obtained by numerically integrating Eq.~(\ref{eq18}) using QuTiP~\cite{qutip_1,qutip_2}. The hole-spin qubit is prepared in \( \lvert 0'\rangle \) [see Fig.~5(a)] with the same EDSR conditions as in Table~\ref{tab:table1}. In the absence of optical coupling, the dynamics reduces to the EDSR results shown in Figs.~4(b) and 4(c). Optical transitions introduce dissipation through the Lindblad jump operators in Eq.~(\ref{eq18}). For example, starting from $\lvert 0'\rangle$, LED photons can excite the carrier to the confined electron states $\lvert 0\rangle$ and $\lvert 1\rangle$ via the $\Gamma_{00'}$ and $\Gamma_{10'}$ absorption channels. The excited states then decay back to the confined hole-state subspace through emission channels $\Gamma_{n'0}$ and $\Gamma_{n'1}$ ($n'=0',1',2',3'$). As EDSR drives population between $\lvert 0'\rangle$ and $\lvert 1'\rangle$, absorption from $\lvert 1'\rangle$ through $\Gamma_{01'}$ and $\Gamma_{11'}$ becomes relevant. Continuous illumination leads to repeated cycles of absorption followed by emission. These repeated absorption--emission cycles act as a persistent noise channel that damps the spin-qubit coherence.
\begin{figure*}
\includegraphics[width=0.80
\textwidth]{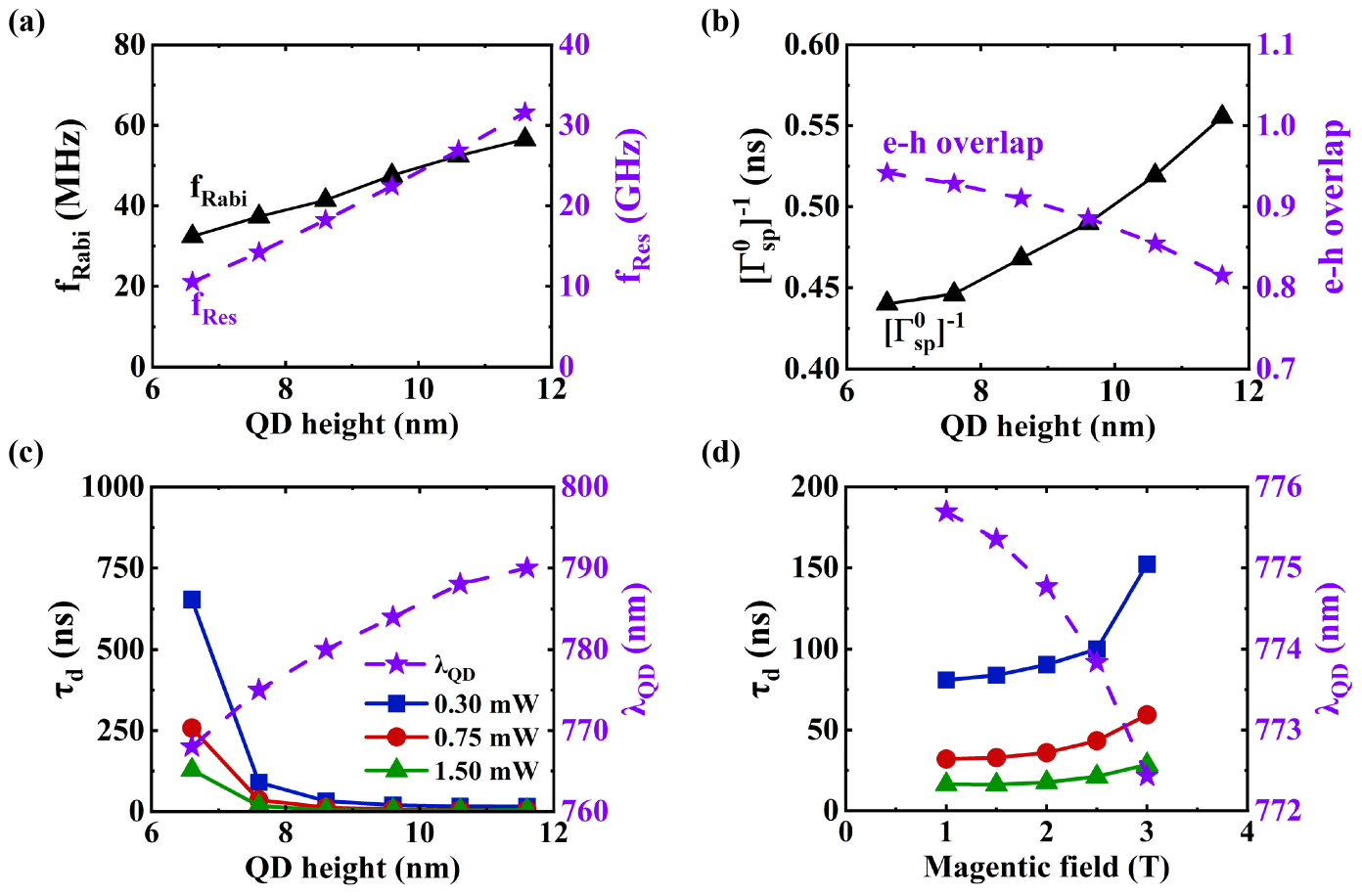}
\caption{\label{fig:Figure_6} For panels (a)--(c), the device parameters are given in Table~\ref{tab:table1}, and $B_0=2~\mathrm{T}$. (a) Hole-spin Rabi frequency (left axis) and EDSR resonance frequency (right axis) versus the height of the SAQD. (b) Spontaneous-emission lifetime, $\left[\Gamma^0_{sp}\right]^{-1}=\left[{\Gamma^0_{0'0}+\Gamma^0_{1'0}}\right]^{-1}$ (left axis), and electron--hole overlap probability, $\lvert\langle\Psi_{n}|\Psi_{n'}\rangle\rvert^2$ (right axis), versus the SAQD height. (c) Photon-induced Rabi-oscillation decay time ($\tau_d$, left axis) and SAQD transition wavelength (right axis) versus SAQD height under illumination by an LED centered at 790~$\mathrm{nm}$. (d) Photon-induced Rabi-oscillation decay time (left axis) and SAQD transition wavelength $\lambda_{\mathrm{QD}}$ versus magnetic field. The symbols and color coding are the same as those in panel (c).}
\end{figure*}
The resulting population dynamics reveal a gradual decay of the EDSR-driven Rabi oscillations, shown in Figs.~5(b) and 5(c). For an optical power of $1.5~\mathrm{mW}$ ($0.3~\mathrm{mW}$), the calculated Rabi-oscillation decay time, $\tau_d$, is approximately $\tau_d \approx 17.5~\mathrm{ns}$ ($90.3~\mathrm{ns}$) at $B_0=2~\mathrm{T}$ and $\tau_d \approx 16.3$$~\mathrm{ns}$ ($80.9~\mathrm{ns}$) at $B_0=1~\mathrm{T}$. The Rabi-oscillation decay time $\tau_d$ is extracted by fitting the occupation probability $O_{0'}(t)$ to the following function,
\begin{equation}
O_{0'}(t)=A\exp\left(-\frac{t}{\tau_d}\right)
\cos\left(\omega_{Rabi} t+\phi\right)+B,
\label{tau_d}
\end{equation}
where $\omega_{Rabi}$ is the Rabi angular frequency, $\phi$ is the initial phase, and $B$ is the steady-state population of the $\lvert 0'\rangle$ state at large times. The oscillation amplitude is constrained by the initial condition $O_{0'}(0)=1$. Since $\langle N_{nn'}\rangle$ is proportional to the LED power, as shown in Eq.~(\ref{eqC4}), reducing the optical power decreases $\langle N_{nn'}\rangle$ and increases $\tau_d$. 

As shown in Tables~\ref{tab:table3} and \ref{tab:table4}, the calculated optical-transition dynamics is dominated by the transition between the lowest $s$-shell confined hole states $\lvert0'\rangle$ and $\lvert1'\rangle$ and electron states $\lvert0\rangle$ and $\lvert1\rangle$, whose spontaneous-emission and absorption rates are substantially larger than those involving the excited hole states $\lvert2'\rangle$ and $\lvert3'\rangle$. Consequently, population leakage into $\lvert2'\rangle$ and $\lvert3'\rangle$ remains negligible and is therefore not shown in Figs.~5(b) and 5(c). This dominance of s-shell optical transitions is consistent with experimental observations in a GaAs SAQD with a similar geometry~\cite{Excitonic_lifetimes}. In addition, because the total emission rate is much larger than the absorption rate under the illumination conditions considered here, carriers rapidly relax back to the confined hole states, leaving the electron-state populations extremely small. At the higher LED power of 1.5~$\mathrm{mW}$, the green curves in Figs.~5(b) and 5(c) exhibit stronger damping of the Rabi oscillations because of the more frequent absorption--emission cycles.

These simulations identify optical power as a critical parameter. While optical excitation is required for spin--photon interfacing, excessive optical power substantially accelerates Rabi-oscillation damping by increasing the rate of interband carrier transitions.

\subsection{Further Analysis and Discussion}

One of the advantages of predictive quantum-device simulation is the ability to evaluate the impact of device geometry and other operational parameters prior to fabrication. Here we examine the dependence of qubit performance on the SAQD height and magnetic field while keeping the remaining device parameters fixed as listed in Table~\ref{tab:table1}.

The EDSR resonance frequency is set by the energy level splitting between the two qubit states $\lvert0'\rangle$ and $\lvert1'\rangle$ [Fig.~5(a)]. One would expect the level splitting, and thus the resonance frequency, to decrease as the SAQD height increases. In contrast, the calculated resonance frequency actually increases with height, as shown by the dashed line (stars) in Fig.~6(a). This unexpected outcome can be understood by the hole Hamiltonian Eq.~(\ref{eq8}): increasing the SAQD height modifies the vertical confinement potential $U_h(\mathbf r)$, thus changing the wave functions and the relative weights of the four components of the envelope function $F_{n', m_J}(\mathbf{r})$. The off-diagonal terms in the Luttinger--Kohn--Foreman Hamiltonian in Eq.~(\ref{eqA1}) couple states with different $m_J$ indices, while the hole Zeeman term in Eq.~(\ref{eq9}) produces the magnetic-field-dependent splitting of the resulting eigenstates. In the SAQD height range studied here, the overall effect of increasing the SAQD height turns out to increase $\lvert E_{1'}-E_{0'}\lvert$, thereby increasing the resonance frequency. As the SAQD height increases further, the EDSR resonance frequency gradually levels off. 

In Fig.~6(a), the calculated Rabi frequency also increases with the height for the range shown. The value of the Rabi frequency is related to the off-diagonal matrix element $\lvert\delta U_{1'0'}(t)\rvert$ of the gate-induced perturbation [Eq.~(\ref{eq12})]. Its height dependence can be understood from the $P$, $Q$, and $S\pm$ parameters of the multiband Luttinger--Kohn--Foreman Hamiltonian in Eqs.~(\ref{eqA1}) and (\ref{eqA2}). Changing the SAQD height modifies the vertical confinement and thus the z-component of the momentum, $k_z$. The $k_z^2$-dependent terms in the $P$ and $Q$ parameters alter the energy separation between the $s$-like and $p$-like states shown in Fig.~3(b). The $k_z$-dependent terms in $S_\pm$ modify the spin--orbit-induced coupling between different $m_J$ components. The static $B_z$ field produces the Zeeman splitting that defines the two qubit states. In general, these effects modify the multiband mixing induced by spin-orbit coupling and, consequently, the magnitude of $\lvert\delta U_{1'0'}(t)\rvert$. Over the height range considered in Fig.~6(a), these combined effects increase $f_{\mathrm{Rabi}}$. We verified numerically that $f_{\mathrm{Rabi}}$ decreases at larger SAQD heights.

Similarly, the cross-band-gap optical transition depends on the SAQD height, as shown in Fig.~6(b). As expected, increasing height reduces the electron--hole overlap measured by $\lvert\langle\Psi_{n}|\Psi_{n'}\rangle\rvert^2$ (dashed line with stars), leading to weaker optical-transition matrix elements (solid line with triangles). Increasing the SAQD height shifts the interband transition wavelength $\lambda_{QD}$, which, for the device and operating parameters considered here, increases the spectral overlap between the LED and the SAQD (Sec.~III-C). As a result, $\tau_d$ decreases drastically as shown in Fig.~6(c). Namely, for a fixed LED spectrum centered at $790~\mathrm{nm}$, increasing the height of the SAQD redshifts the dominant interband transition wavelength $\lambda_{QD}$ toward $790~\mathrm{nm}$, leading to higher rates of repeated absorption--emission cycles. Consequently, despite reduced electron--hole overlap and weaker intrinsic optical-transition strength, the Rabi oscillations decay faster as the SAQD height increases. This effect is most pronounced near a SAQD height of approximately $11.6~\mathrm{nm}$. At this height, the dominant transition wavelength $\lambda_{QD}$ coincides with the central LED  wavelength of $790~\mathrm{nm}$ at which the LED spectral density is maximal. These results suggest that SAQD geometry and the optical spectrum should be considered together when evaluating photon-induced damping of Rabi oscillations in spin--photon interfaces.

Fig.~6(d) shows that the Rabi oscillation decay time $\tau_d$ increases with the strength of the external magnetic field [$B_0$ of Eq.~(1)]. This trend originates largely from Zeeman splitting, which alters the interband transition energy and consequently changes the spectral overlap between the SAQD transitions and the fixed LED spectrum centered at $790~\mathrm{nm}$. Over the magnetic-field range in Fig.~6(d), the dominant SAQD transition moves farther away from the center of the LED spectrum, which reduces the absorption rate and the number of repeated absorption--emission cycles. Since these optical cycles are the source of photon-induced damping, the calculated decay of the Rabi oscillations is slower at higher magnetic fields. The magnetic field strength introduces another operational trade-off. Larger magnetic fields improve spin controllability by increasing the Rabi frequency, but they also modify the optical-transition landscape in spin--photon coupling.

The results suggest several practical considerations for SAQD-based spin--photon interfaces. First, for our simulation conditions, increasing the height of the SAQD enhances the Rabi frequency by strengthening the gate-induced coupling of the confined hole states. Second, increasing the SAQD height reduces the electron--hole wave function overlap, resulting in weaker optical-transition matrix elements and longer spontaneous-emission lifetimes. Third, photon-induced Rabi-oscillation damping is critically governed by the spectral overlap between the SAQD transitions and the external light source. Even small changes in this overlap can produce large variations in the Rabi-oscillation damping. Fourth, reducing the optical power increases the Rabi-oscillation decay time by suppressing photon-induced carrier excitation. Finally, the magnetic-field strength affects spin–photon interfaces by tuning the optical transitions. Collectively, these results demonstrate that quantum-device simulation provides a useful means of identifying potentially optimal operating conditions for spin–photon interface hardware before fabrication.

\section{Summary}

We have developed a quantum-device simulation framework for quantitatively modeling spin--photon interactions, and have applied it to predict the optical-transition-induced damping of Rabi oscillations in SAQD spin qubits. The approach integrates multiband electronic-structure calculations, electric-dipole spin resonance, optical-transition modeling, and Lindblad master-equation dynamics within a single computational workflow. By combining device physics with open-quantum-system simulations, the framework establishes a quantitative connection between SAQD geometry, optical excitation conditions, and an important metric of qubit performance. 

Application of the framework to GaAs SAQD spin qubits demonstrates that photon-induced Rabi-oscillation damping strongly depends on the spectral overlap between the optical source and SAQD transitions. The systematic analysis of device geometry, magnetic field strength, optical power, and SAQD transition properties identifies the key factors governing the operation of spin--photon interfaces.

More broadly, this work advances modeling methodologies for emerging quantum technologies. Similar to the role of simulation-driven design in classical semiconductor engineering, integrated quantum-device simulation frameworks can accelerate the development of scalable quantum hardware by enabling rapid exploration of design spaces prior to fabrication. The framework presented here is readily extendable to different material platforms, decoherence mechanisms, and quantum-photonic architectures, including cavity-coupled systems, spin--photon entanglement devices, quantum repeaters, and distributed quantum-information networks. We anticipate that integrated quantum-device simulation for spin–photon systems will become a key enabling technology for the predictive design, co-optimization, and scaling of future quantum-device and quantum-information architectures.

\begin{acknowledgments}
J.J. and H.G. thank Prof. Thomas Szkopek for discussions on the physics of quantum repeaters, Dr. Marek Korkusinski for sharing valuable insights into QD--photon interactions, and Prof. David Cooke for discussions on the LED--SAQD spectral overlap analysis. We gratefully acknowledge financial support from the Natural Sciences and Engineering Research Council of Canada (NSERC) and Photonique Quantique Quebec (PQ2). This work benefits from the co-authors’ RQMP membership https://doi.org/10.69777/309032. We thank the Digital Research Alliance of Canada for computational facilities that made the numerical modeling possible.
\end{acknowledgments}

\appendix
\section{The Luttinger--Kohn--Foreman Hamiltonian}
\label{appendix:A}
In this Appendix, we summarize the Luttinger--Kohn—Foreman Hamiltonian for holes. The corresponding four-band model describes the effective Schrödinger equation for confined holes~\cite{Analysis_and3D_TCAD,Robust},
\begingroup
\renewcommand{\arraystretch}{1} 
\setlength{\arraycolsep}{0.5pt}     
\begin{equation}
\begin{pmatrix}
P+Q & 0 & -S_- & R \\
0 & P+Q & -R^\dagger & -S_+ \\
-S_-^\dagger & -R & P-Q & C \\
R^\dagger & -S_+^\dagger & C^\dagger & P-Q
\end{pmatrix}
\begin{pmatrix}
F_{n', \frac{3}{2}} \\
F_{n', -\frac{3}{2}} \\
F_{n', \frac{1}{2}} \\
F_{n', -\frac{1}{2}}
\end{pmatrix}
=E
\begin{pmatrix}
F_{n', \frac{3}{2}} \\
F_{n', -\frac{3}{2}} \\
F_{n', \frac{1}{2}} \\
F_{n', -\frac{1}{2}}
\end{pmatrix}
\label{eqA1}
\end{equation}
\endgroup
where the left-hand-side $4\times4$ matrix is the Luttinger--Kohn--Foreman Hamiltonian corresponding to the $D$ matrix introduced in Sec.~II, with
\begin{equation}
\begin{aligned}
&P = E_V(\mathbf{r}) + \frac{\hbar^2}{2m_0} \left( k_x \gamma_1 k_x + k_y \gamma_1 k_y + k_z \gamma_1 k_z \right), \\
&Q = \frac{\hbar^2}{2m_0} \left( k_x \gamma_2 k_x + k_y \gamma_2 k_y - 2 k_z \gamma_2 k_z \right), \\
&R = -\frac{\hbar^2 \sqrt{3}}{2m_0} k_- \bar{\gamma} k_- + \frac{\hbar^2 \sqrt{3}}{2m_0} k_+ \mu k_+, \\
&S_\pm = \frac{\hbar^2 \sqrt{3}}{m_0} \left[ k_\pm (\sigma - \delta) k_z + k_z \pi k_\pm \right], \\
&C = \frac{\hbar^2}{m_0} \left[ k_z (\sigma - \delta - \pi) k_- - k_- (\sigma - \delta - \pi) k_z \right].\\
\end{aligned}
\label{eqA2}
\end{equation}
In Eq.~(A2), $E_V$ is the energy of the valence band after integrating out the cell-periodic Bloch functions; $\hbar$ is the reduced Planck constant; $k_{x}$, $k_{y}$, and $k_{z}$ are the wave-vector components along the three Cartesian axes. When a magnetic field is applied, the wave vector is transformed as $\mathbf{k} \rightarrow \mathbf{k} - \frac{e}{\hbar} \mathbf{A}$. 
The Luttinger--Kohn parameters $\gamma_1=6.98$, $\gamma_2=2.06$, and $\gamma_3=2.93$ characterize GaAs~\cite{Giant_Zeeman}. The remaining quantities in Eq.~(\ref{eqA2}) are defined by the following relations,
\begin{equation}
\begin{aligned}
k_{\pm} &= k_x \pm i k_y, & k_{\parallel}^2 &= k_x^2 + k_y^2, \\
\bar{\gamma} &= \frac{1}{2}(\gamma_3 + \gamma_2), & \mu &= \frac{1}{2}(\gamma_3 - \gamma_2), \\
\sigma &= \bar{\gamma} - \frac{1}{2}\delta, & \pi &= \mu + \frac{3}{2}\delta, \\
\delta &= \frac{1}{9}(1 + \gamma_1 + \gamma_2 - 3\gamma_3)
\end{aligned}
\label{eqA3}
\end{equation}
Because the Luttinger--Kohn--Foreman Hamiltonian [Eq.~(\ref{eqA1})] contains off-diagonal coupling terms among heavy and light holes, the resulting eigenstates are superpositions of basis states with different spin components.

\section{Spontaneous-Emission Rate}
\label{Appendix:B}

In this Appendix, we summarize the expressions used to calculate spontaneous-emission rates between confined electron and hole states. Spontaneous emission arises from the coupling of the SAQD to external electromagnetic fields. After evaluating the photon-state part of the light-matter interaction, the squared effective interband optical matrix element entering Eq.~(\ref{eq17}) is written as~\cite{Quantum_Optics}:
\begin{equation}
\begin{aligned}
 \left\lvert  {\left\langle {{\Psi _{n'}}} \right\lvert \hat H_{\mathrm{int}}\left\lvert  {{\Psi_{n}}} \right\rangle } \right\lvert ^2 &=\frac{e^2\hbar}
{2m_0^2\epsilon_0\epsilon_r\omega \mathcal{V}}
\left\lvert  \left\langle \Psi_{n'} \left\lvert  \hat{\mathbf e}_{\lambda} \cdot \hat{\mathbf p} \right\lvert  \Psi_n \right\rangle \right\lvert ^2 \\
&=\frac{e^2 \hbar}{2m^2_0 \epsilon_{0} \epsilon_{r} \omega \mathcal{V}} \left\lvert M_{T,n'n}\right\lvert ^2,
\end{aligned}
\label{eqB1}
\end{equation}
where $\hat{\mathbf{e}}_\lambda$ is the photon-polarization unit vector and $\hat{\mathbf{p}}$ is the momentum operator; $\epsilon_0$ ($\epsilon_r=n_r^2$) is the vacuum (relative) permittivity of the material; $\mathcal{V}$ is the photon quantization volume; $\omega$ is the angular frequency of the photon mode. $M_{T,n'n}$ denotes the interband momentum matrix element associated with the polarization vector ${\hat{\mathbf e}}_\lambda$. The integration in Eq.~(\ref{eq16}) is understood to include the sum over the two transverse photon polarizations. The confined wave functions are expanded in terms of cell-periodic Bloch functions and slowly varying envelope functions. Because envelope functions vary negligibly over a single unit cell, the interband momentum matrix element can be separated into cell-periodic Bloch-function matrix elements and envelope function overlap integrals~\cite{spin_orbit_coupling},
\begin{widetext}
\begin{equation}
\begin{aligned}
\lvert M_{T,n'n}\lvert ^2 = & \Bigg\lvert \Bigg[ 
   \left\langle u_{\tfrac{1}{2},\tfrac{1}{2}} \middle\lvert  \hat{e}_\lambda \cdot \hat{p} \middle\lvert  u_{\tfrac{3}{2},\tfrac{3}{2}} \right\rangle \left\langle F_{n, \tfrac{1}{2}} \middle\lvert  F_{n', \tfrac{3}{2}} \right\rangle  +   \left\langle u_{\tfrac{1}{2},-\tfrac{1}{2}} \middle\lvert  \hat{e}_\lambda \cdot \hat{p} \middle\lvert  u_{ \tfrac{3}{2},-\tfrac{3}{2}} \right\rangle \left\langle F_{n, -\tfrac{1}{2}} \middle\lvert  F_{n', -\tfrac{3}{2}} \right\rangle  \\
& +  \left\langle u_{\tfrac{1}{2},\tfrac{1}{2}} \middle\lvert  \hat{e}_\lambda \cdot \hat{p} \middle\lvert  u_{\tfrac{3}{2},-\tfrac{1}{2}} \right\rangle \left\langle F_{n, \tfrac{1}{2}} \middle\lvert  F_{n', -\tfrac{1}{2}} \right\rangle  + \left\langle u_{\tfrac{1}{2},-\tfrac{1}{2}} \middle\lvert  \hat{e}_\lambda \cdot \hat{p} \middle\lvert  u_{\tfrac{3}{2},\tfrac{1}{2}} \right\rangle \left\langle F_{n, -\tfrac{1}{2}} \middle\lvert  F_{n', \tfrac{1}{2}} \right\rangle\\
& + \left\langle u_{\tfrac{1}{2},-\tfrac{1}{2}} \middle\lvert  \hat{e}_\lambda \cdot \hat{p} \middle\lvert  u_{\tfrac{3}{2},-\tfrac{1}{2}} \right\rangle \left\langle F_{n, -\tfrac{1}{2}} \middle\lvert  F_{n', -\tfrac{1}{2}} \right\rangle+  \left\langle u_{\tfrac{1}{2},\tfrac{1}{2}} \middle\lvert  \hat{e}_\lambda \cdot \hat{p} \middle\lvert  u_{\tfrac{3}{2},\tfrac{1}{2}} \right\rangle \left\langle F_{n, \tfrac{1}{2}} \middle\lvert  F_{n', \tfrac{1}{2}} \right\rangle  \Bigg] \Bigg\lvert ^2.
\end{aligned}
\label{eqB2}
\end{equation}
\end{widetext}
The unit-cell-periodic conduction-band and valence-band Bloch basis functions ($u_{1/2, m_J}$ and $u_{3/2, m_{J}}$) have predominantly $s$-like and $p$-like orbital character, respectively, giving rise to a nonzero interband momentum matrix element. For the cell-periodic part, the matrix elements of the momentum operator between $u_{1/2,m_J}$ and $u_{3/2,m_J}$ can be derived and expressed in terms of the interband momentum matrix element $M$.
We finally obtain the spontaneous-emission rate,
\begin{widetext}
\begin{equation}
\begin{aligned}
\Gamma^0_{n'n} = C^0_{n'n} &\Bigg[ 
 \frac{1}{2} \left\lvert  \left\langle F_{n, \tfrac{1}{2}} \middle\lvert  F_{n', \tfrac{3}{2}} \right\rangle \right\lvert ^2 
+ \frac{1}{2} \left\lvert  \left\langle F_{n, -\tfrac{1}{2}} \middle\lvert  F_{n', -\tfrac{3}{2}} \right\rangle \right\lvert ^2 \\
& + \frac{1}{6} \left\lvert  \left\langle F_{n, \tfrac{1}{2}} \middle\lvert  F_{n', -\tfrac{1}{2}} \right\rangle \right\lvert ^2
+ \frac{1}{6} \left\lvert  \left\langle F_{n, -\tfrac{1}{2}} \middle\lvert  F_{n', \tfrac{1}{2}} \right\rangle \right\lvert ^2\\
& + \frac{1}{3} \left\lvert  \left\langle F_{n, -\tfrac{1}{2}} \middle\lvert  F_{n', -\tfrac{1}{2}} \right\rangle \right\lvert ^2
+ \frac{1}{3} \left\lvert  \left\langle F_{n, \tfrac{1}{2}} \middle\lvert  F_{n', \tfrac{1}{2}} \right\rangle \right\lvert ^2\Bigg]
\end{aligned}
\label{eqB3}
\end{equation}
\end{widetext}
where $C^0_{n'n} = \frac{{{n_{r} e{^2}}{\hbar \omega_{n'n}}}}{{6 \pi \hbar^2 {m_0}{c^3}{\varepsilon _{0}}}}{E_p}$ is a transition-dependent prefactor; $n_r=3.6$ is the refractive index of GaAs; $e$ is the electron charge; $c$ is the speed of light; $\hbar\omega_{n'n}=E_{n}-E_{n'}$ is the transition energy between the states $\Psi_{n}$ and $\Psi_{n'}$; $E_p=\frac{2M^2}{m_0}=28.8~\mathrm{eV}$ is the Kane energy~\cite{coldren2012diode} associated with the interband matrix element of the cell-periodic Bloch functions for GaAs; $\varepsilon_0$ is the permittivity of free space. In deriving from Eq.~(\ref{eqB2}) to Eq.~(\ref{eqB3}), all cross terms vanish after integration over the solid angle and application of the optical selection rules.

\section{Mean Number of Photons}
\label{Appendix:C}
In this Appendix, we estimate the mean photon number by considering spectral overlap between the LED source and the SAQD transition linewidths. We assume that the LED source has a Gaussian spectral profile, as described in the main text. From the spectral power density $D_{LED}(\nu)$ of the LED, the flux density entering the SAQD can be written as
\begin{equation}
\Phi(\nu)=\frac{P_{total}}{h\nu_{QD,n'n}}D_{LED}(\nu)
\label{eqC1}
\end{equation}
where $P_{total}$ denotes the optical power reaching the SAQD. We approximate the photon energy by $h\nu_{QD,n'n}$ over the relevant spectral-overlap region.

For the SAQD, we consider spectral broadening due to the optical transitions. The resonance frequency of the SAQD is denoted by $\nu_{QD}$, and the spectral linewidth is given by $\Delta \nu_{QD, n'n}=\frac{1}{2\pi} \Gamma^0_{n'n}$. The SAQD absorption cross section $A_{QD}$ is modeled by a Lorentzian profile corresponding to a lifetime-broadened transition centered at the SAQD resonance frequency,
\begin{equation}
A_{QD,n'n}(\nu)=A_0\frac{\left(\frac{\Delta \nu_{QD,n'n}}{2}\right)^2}{\left(\nu-\nu_{QD,n'n}\right)^2+\left(\frac{\Delta \nu_{QD,n'n}}{2}\right)^2}
\label{eqC2}
\end{equation}
where $A_0=\frac{c^2}{2\pi (\nu_{QD,n'n}n_r)^2}$ is the peak cross section, derived from the SAQD absorption rate with the Einstein A and B coefficients. The absorption rate depends on the spectral overlap between the Lorentzian SAQD response and the frequency-dependent photon flux of the LED source by integrating over frequency,
\begin{equation}
\Gamma_{nn'}= \Gamma^0_{n'n}\langle N_{nn'}\rangle=\int_0^{\infty}\Phi(\nu)\frac{A_{QD,n'n}(\nu)}{A_{beam}}d\nu
\label{eqC3}
\end{equation}
where $[\Gamma^0_{n'n}]^{-1}$ is the spontaneous-emission lifetime associated with the transition between the $n$ and $n'$ states. Since the spectral width of the LED is much broader than the natural linewidth of the SAQD transition, the Lorentzian SAQD response in Eq.~(\ref{eqC2}) can be approximated by a delta function, $A_0 \pi \left(\frac{\Delta \nu_{QD}}{2}\right)\delta(\nu-\nu_{QD})$. Substituting this approximation into Eq.~(\ref{eqC3}), the mean number of photons is given by
\begin{equation}
\begin{aligned}
\langle N_{nn'}\rangle=&\frac{1}{2}\sqrt{\frac{ln2}{\pi}}\frac{P_{total}}{h\nu_{QD,n'n}}\frac{A_0}{A_{beam}}\frac{1}{\Delta\nu_{LED}} \\
&\times\text{exp}\left[-\frac{(\nu_{QD,n'n}-\nu_{LED})^2}{2\sigma_{LED}^2} \right].
\label{eqC4}
\end{aligned}
\end{equation}

\bibliography{apssamp}

\end{document}